\documentclass[reprint,superscriptaddress,floatfix,longbibliography]{revtex4-2}
\usepackage{blindtext}
\usepackage{amsfonts}
\usepackage{amsmath}
\usepackage{amssymb}
\usepackage{graphicx}%
\usepackage{color}
\usepackage{epstopdf}
\usepackage {multirow}

\usepackage[capitalize]{cleveref}

\usepackage[LGR,T1]{fontenc}
\usepackage[latin9]{inputenc}
\usepackage{array}
\usepackage{booktabs}
\usepackage{textcomp}
\usepackage{multirow}
\usepackage{diagbox}
\usepackage{braket}
\usepackage{url}
\makeatletter

\ProvideTextCommand{\~}{LGR}[1]{\char126#1}

\makeatother

\DeclareSymbolFont{myletters}{OML}{ztmcm}{m}{it}
\DeclareMathSymbol{\uplambda}{\mathord}{myletters}{"15}
\begin{document}
	\title{Symmetry Analysis of Magnetoelectric Coupling Effect in All Point Groups}
	\author{Xinhai Tu}
	\affiliation{National Laboratory of Solid State Microstructures and School of Physics, Nanjing University, Nanjing 210093, China}
	\affiliation{Collaborative Innovation Center of Advanced Microstructures, Nanjing University, Nanjing 210093, China}
	\author{Di Wang}
	\affiliation{National Laboratory of Solid State Microstructures and School of Physics, Nanjing University, Nanjing 210093, China}
	\affiliation{Collaborative Innovation Center of Advanced Microstructures, Nanjing University, Nanjing 210093, China}
		\author{Hanjing Zhou}
	\affiliation{National Laboratory of Solid State Microstructures and School of Physics, Nanjing University, Nanjing 210093, China}
	\affiliation{Collaborative Innovation Center of Advanced Microstructures, Nanjing University, Nanjing 210093, China}
	\author{Songsong Yan}
	\affiliation{National Laboratory of Solid State Microstructures and School of Physics, Nanjing University, Nanjing 210093, China}
	\affiliation{Collaborative Innovation Center of Advanced Microstructures, Nanjing University, Nanjing 210093, China}
	\author{Huimei Liu}
	\affiliation{National Laboratory of Solid State Microstructures and School of Physics, Nanjing University, Nanjing 210093, China}
	\affiliation{Collaborative Innovation Center of Advanced Microstructures, Nanjing University, Nanjing 210093, China}
	\author{Hongjun Xiang}
	\affiliation{Key Laboratory of Computational Physical Sciences (Ministry of Education), Institute of Computational Physical Sciences, and Department of Physics, Fudan University, Shanghai 200433, China}
	\author{Xiangang Wan}
	\email{xgwan@nju.edu.cn}
	\affiliation{National Laboratory of Solid State Microstructures and School of Physics, Nanjing University, Nanjing 210093, China}
	\affiliation{Collaborative Innovation Center of Advanced Microstructures, Nanjing University, Nanjing 210093, China}
	\affiliation{Hefei National Laboratory, Hefei 230088,China}
	\affiliation{Jiangsu Physical Science Research Center, Nanjing University, Nanjing 210093, China}
	\begin{abstract}
Symmetry analysis provides crucial insights into the magnetoelectric coupling effect in type-II multiferroics. In this Letter, we comprehensively investigate couplings between electric polarization and inhomogeneous magnetization across all 32 crystallographic point groups using a phenomenological Landau theory. Our theory successfully explains the ferroelectric polarizations in all known type-II multiferroics characterized by incommensurate magnetic orders. In addition, we predict 12 promising type-II multiferroic candidates with the highest magnetic transition temperature of 84 K through systematic screening of MAGNDATA database. Furthermore, we find that the collinear spin-sinusoidal texture emerges as a previously unrecognized source of ferroelectric polarization. We also demonstrate that topological ferroelectric vortex states can be induced by ferromagnetic vortex configurations in uniaxial point groups, opening a route to realizing coexisting multiple-vortex states in multiferroics.
	\end{abstract}
	\maketitle
\textit{Introduction--}Intrinsic magnetoelectric coupling (MEC) effects in type-II multiferroic materials \cite{Tokura_2014,Cheong2007,10.1093/nsr/nwz023,Dong2015,Wang2009,RR,Mostovoy2024,Gao2024,Song2022,Han2023,Du2022,PhysRevLett.132.086802,https://doi.org/10.1002/anie.201609762,PhysRevB.91.214415,Perks2012,PhysRevB.86.060403,PhysRevLett.98.267205,PhysRevLett.101.067204,PhysRevLett.101.067204,PhysRevB.73.220401,PhysRevB.87.014429,PhysRevLett.106.167206,Kitagawa2010,Murakawa2008,PhysRevLett.94.137201,Ishiwata2008,PhysRevLett.96.207204,Kim2012,PhysRevLett.111.017202,Newnham1978,Kimura2008,https://doi.org/10.1002/adma.201200734,PhysRevB.82.064424,PhysRevB.77.144101,PhysRevLett.98.057601,PhysRevLett.100.127201,PhysRevLett.97.097203,PhysRevB.74.184431,PhysRevLett.95.087205,PhysRevB.76.184418,PhysRevLett.92.257201,PhysRevLett.92.257201} have garnered significant attention due to their fundamental physical implications and potential applications. In particular, the coexistence of ferromagnetic and ferroelectric vortex states \cite{Chen2020,2016,SnchezSantolino2024,Donnelly2020} in multiferroics is promising for the development of next-generation multifunctional devices. Various microscopic models \cite{PhysRevLett.95.057205,PhysRevB.73.094434,PhysRevB.83.174432,PhysRevB.74.224444,PhysRevB.76.144424,PhysRevLett.107.157202,PhysRevLett.97.227204,PhysRevLett.100.077202,PhysRevB.73.184433,2501.05025,PhysRevLett.134.066801,PhysRevLett.101.037209} have been proposed to explain prominent MEC effects in type-II multiferroics. However, identifying the predominant microscopic mechanism in specific materials remains a significant challenge.

In contrast to microscopic approaches, symmetry analysis \cite{Schmid2008,Cheong2022,PhysRevB.109.104413} not only provides a universal and powerful framework for elucidating the MEC phenomena in intricate systems but also potentially aids in understanding the underlying microscopic origins of MEC effects. The landmark advancement was Mostovoy's phenomenologically study of MECs in the centrosymmetric cubic system, utilizing a continuum field theory of the Ginzburg-Landau type \cite{PhysRevLett.96.067601}. This study successfully described the ferroelectricity introduced by incommensurate cycloidal spin density wave states in TbMnO$_{3}$. However, it is worth noting that TbMnO$_{3}$ crystallizes in an orthorhombic rather than cubic system, making it inappropriate for the direct applicability of Mostovoy's theory. Recently, Mostovoy's theory has also been employed to explain MEC effects in two-dimensional material NiI$_{2}$ \cite{Amini,Fumega2022,Wu2024}, which possesses a non-cubic lattice structure as well. Moreover, Mostovoy's theory predominantly addresses polarization resulting from specific magnetic orders, such as noncollinear cycloidal order. This raises an essential question: can other magnetic configurations, including noncollinear screw orders, collinear sinusoidal orders, or topological ferromagnetic vortices, also couple effectively to ferroelectric polarization? Specifically, ferromagnetic vortices, typically exhibit a much larger size than ferroelectric vortices, making their simultaneous presence within a single multiferroic material challenging \cite{Zheng2017}. Nonetheless, the intriguing possibility remains that ferroelectric vortices could be induced by ferromagnetic vortices, potentially enabling multi-vortex states to coexist. Lately, although Tanygin \cite{TANYGIN20121878} extended the phenomenological theory to tetragonal and orthorhombic crystal systems, his treatment remains incomplete and imprecisely. Other symmetry-based studies have either been narrowly tailored to specific materials \cite{Ribeiro2010,PhysRevB.90.214427} or have employed overly intricate theoretical methods \cite{PhysRevB.83.174432,Matsumoto2017}, which may pose challenges for broader adoption or intuitive interpretation. Thus, it is crucial and pressing to develop a comprehensive theory for MEC effects to address these issues.

In this Letter, we present a systematical investigation of MECs between electric polarization $\mathbf{P}$ and inhomogeneous magnetization $\mathbf{M}$ across all 32 point groups by a continuum field theory. Our theory explains successfully the ferroelectric polarizations in all known type-II multiferroics with incommensurate magnetic orders. Through systematic screening of MAGNDATA database \cite{Gallego2016,https://doi.org/10.1107/S1600576716015491}, we identify 12 candidate materials by our theory with a maximum magnetic transition temperature of 84 K. We also find that the type-I multiferroics can simultaneously exhibit the characteristics of type-II multiferroics. In addition, our theory predicts that the collinear spin-sinusoidal texture is one of the sources of ferroelectric polarization besides the conventional noncollinear spin-cycloidal and spin-screwed magnetic orders (Fig. 1). The direction of induced polarization depends on the specific crystal symmetry of multiferroics. Furthermore, we find that the topological ferroelectric vortex states can be driven by ferromagnetic vortex states in 11 uniaxial point groups $C_{6h}$, $C_{3h}$, $C_{6}$, $C_{4h}$, $S_{4}$, $C_{4}$, $C_{3i}$, $C_{3}$, $C_{2h}$, $C_{s}$, and $C_{2}$, establishing a novel pathway toward achieving the coexistence of multiple-vortex states and avoiding the problem of size mismatch.
\begin{table*}[ptb]
	\caption{Experimentally confirmed data of 5 representative type-II multiferroic materials: the point groups, the magnetic orders, the propagation vectors, the spin spiral axes, and the polarization axes given in the Cartesian coordinate system  (see SM). The last two columns exhibit systematic comparisons of polarizations of our work and Mostovoy's study. Symbol * indicates that the agreement between Mostovoy's result and experimental data appears to be fortuitous.}
	\begin{tabular}{|c|c|c|c|c|c|c|c|c|c|}
		\hline
		Type-II multiferroics&Point group&Magnetic order&Propagation Vector&Spin rotation axis&Polarization axis&Mostovoy&Our work\\
		\hline
		TbMnO$_{3}$ \cite{PhysRevLett.95.087206,PhysRevLett.92.257201} &$D_{2h}$&Cycloidal&(0,1,0)&(1,0,0)&(0,0,1)&(0,0,1)*&(0,0,1)\\
		CoCr$_{2}$O$_{4}$ \cite{PhysRevLett.96.207204}&$O_{h}$&Cycloidal&(-1,1,0)&(0,0,1)&(1,1,0)&(1,1,0)&(1,1,0)\\
		SrMnGe$_{2}$O$_{6}$ \cite{PhysRevB.101.235109}&$C_{2h}$&Cycloidal&(0,0,1)&(1,0,0)&in plane&(0,1,0)&in plane\\
		CuFeO$_{2}$ \cite{PhysRevB.73.220401}&$D_{3d}$&Screw&(1,-1,0)&(1,-1,0)&(1,-1,0)&-&(1,-1,0)\\
		Cu$_{2}$OSeO$_{3}$ \cite{PhysRevB.86.060403}&$T$&Screw&(1,1,1)&(1,1,1)&(1,1,1)&-&(1,1,1)\\
		&$T$&Screw&(1,1,0)&(1,1,0)&(0,0,1)&-&(0,0,1)\\
		\hline
	\end{tabular}
\end{table*}
\begin{figure}
	\centering
	\includegraphics[width=8cm]{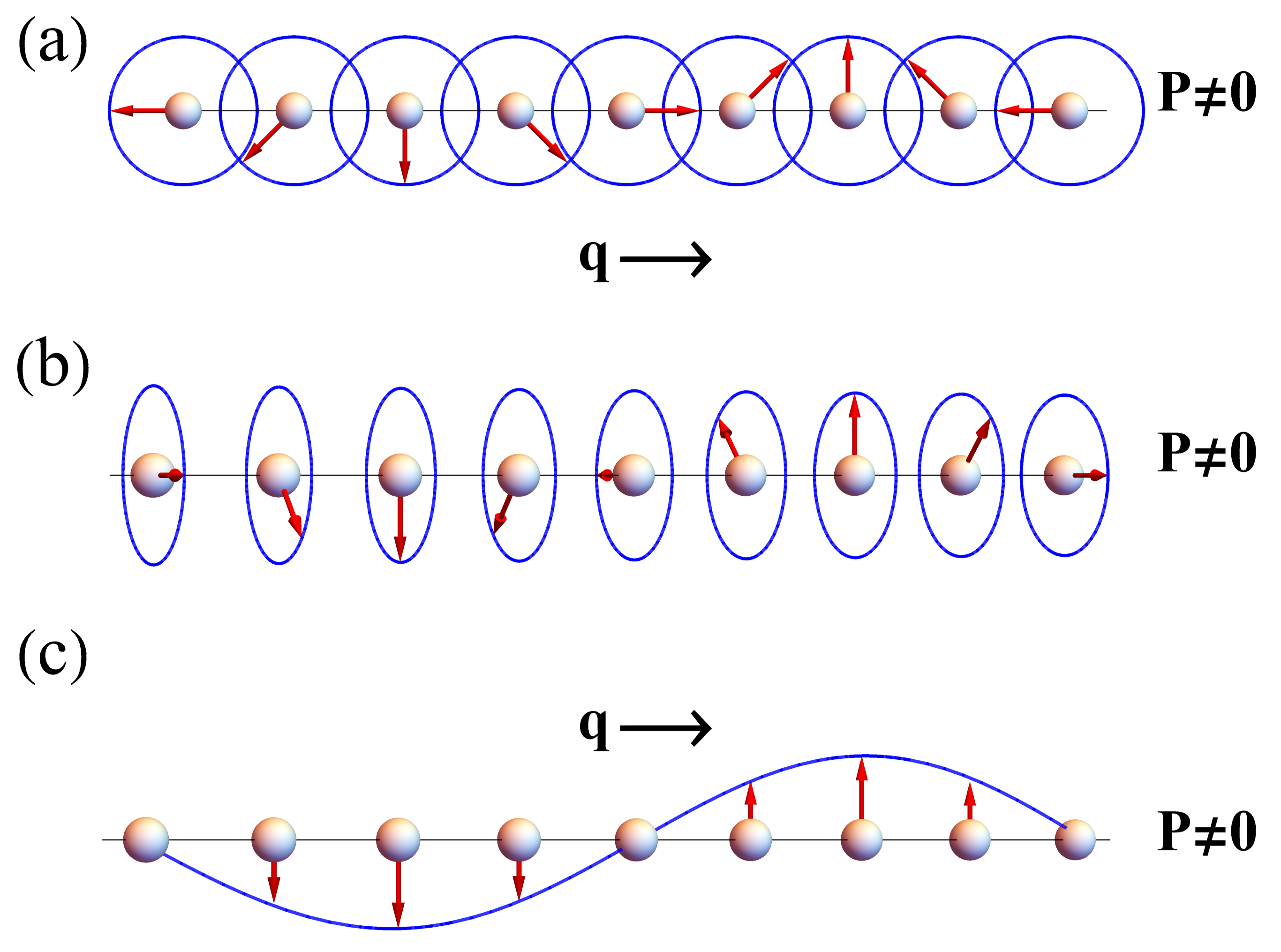}
	\caption{Schematic illustrations of types of spiral magnetic structures. (a) Cycloidal, (b) screwed, and (c) sinusoidal spin textures can generate nonzero ferroelectric polarizations along any direction depending on the specific crystal symmetry.}\label{fig1}
\end{figure}

\textit{Continuum field theory--}Incommensurate magnetic structures are largely insensitive to details of crystal structure and can be properly described by a continuum field theory. The thermodynamic potential depends on the electric polarization $\mathbf{P}$, the magnetization $\mathbf{M}$, and the derivative of magnetization $\nabla\mathbf M$ in our case. The coupling between electric polarization and inhomogeneous magnetization is restricted by crystallographic symmetry. The time-reversal symmetry (TRS) requires the lowest-order coupling to be quadratic in $\mathbf M$. Thus, the MEC terms linear and quadratic in the first derivatives of $\mathbf M$ take the forms $\alpha P_{a}M_{b}\nabla_{c} M_{d}$ and $\beta P_{a}\nabla_{b}M_{c}\nabla_{d} M_{e}$, respectively. The subscripts represent different spatial components $xyz$, and $\alpha$ and $\beta$ are the coupling coefficients, which rely on the material itself. The linear term inherently preserves both TRS and spatial inversion symmetry (SIS) itself, making it symmetry-allowed in all point groups. In contrast, the quadratic term violates SIS and only allowed in the point groups lacking SIS: $C_{2}$, $C_{s}$, $D_{2}$, $C_{2v}$, $C_{4}$, $S_{4}$, $D_{4}$, $C_{4v}$, $D_{2d}$, $C_{3}$, $D_{3}$, $C_{3v}$, $C_{6}$, $C_{3h}$, $D_{6}$, $C_{6v}$, $D_{3h}$, $T$, $O$, and $T_{d}$. Since the invariance principle of the thermodynamic potential under coordinate transformations and group operations, the coupling coefficients $\alpha$ and $\beta$ must be even- and odd-parity, respectively, under SIS. It is notable that symmetric terms of the form $\alpha P_{a}M_{b}\nabla_{c} M_{d}+\alpha P_{a}M_{d}\nabla_{c} M_{b}$ and $\beta P_{a}\nabla_{b}M_{c}\nabla_{d} M_{e}+\beta P_{a}M_{c}\nabla_{b}\nabla_{d} M_{e}$ are rigorously excluded through integration by parts, as they reduce to boundary terms with vanishing contributions to the uniform polarization. Crucially, the equivalence between $\beta P_{a}\nabla_{b}M_{c}\nabla_{d} M_{e}-\beta P_{a}M_{e}\nabla_{b}\nabla_{d} M_{c}$ and $\beta P_{a}\nabla_{d}M_{c}\nabla_{b} M_{e}-\beta P_{a}M_{c}\nabla_{b}\nabla_{d} M_{e}$ is identified by a constant energy shift under full spatial integration through the whole crystal.
\begin{table*}[ptb]
	\caption{The point groups, the magnetic orders, the propagation vectors, the spin rotation axes, and the magnetic transition temperatures of 12 theoretically predicted type-II multiferroic candidates are given in the Cartesian coordinate system (see SM). The calculated polarization axes are given in the last column.}
	\begin{tabular}{|l|c|c|c|c|c|c|c|}
		\hline
		Type-II multiferroics&Point group&Magnetic order&Propagation vector&Spin rotation axis&Transition temperature (K)&Polarization axis\\
		\hline
		1. CrB$_{2}$&$D_{6h}$&Cycloidal \cite{Funahashi1977}&(1,0,0)&(0,1,0)&84&(0,0,1)\\
		2. LiFeAs$_{2}$O$_{7}$&$C_{2}$&Helical-cycloidal \cite{PhysRevB.88.214433}&(1,0,0)&see SM&35&see SM\\
		3. VPO$_{4}$&$D_{2h}$&Cycloidal \cite{Glaum1996}&(1,0,0)&(0,0,1)&25.5&(0,1,0)\\
		4. K$_{2}$MnS$_{2}$&$D_{2h}$&Cycloidal \cite{PhysRevMaterials.3.064404}&(1,0,0)&(0,0,1)&17&(0,1,0)\\
		5. RbMnBr$_{3}$&$D_{6h}$&Cycloidal \cite{10.1063/1.2946987}&(1,1,0)&(0,0,1)&8.8&(1,-1,0)\\
		6. MnSO$_{4}$&$D_{2h}$&Cycloidal \cite{PhysRev.140.A2139}&(1,0,0)&(0,0,1)&7.2 &(0,1,0)\\
		7. GdCrO$_{3}$&$D_{2h}$&Cycloidal \cite{Manuel2025}&(0.08,0.32,0)&(0,1,0)&2.4&(0,0,1)\\
		8. Ni$_{2}$InSbO$_{6}$&$C_{3}$&Screw \cite{Ivanov2013}&(0,1,0)&(0,1,0)&76&see SM\\
		9. Ni$_{2}$ScSbO$_{6}$&$C_{3}$&Screw \cite{Ivanov2013}&(0,1,0)&(0,1,0)&60&see SM\\
		10. MnBi$_{2}$Se$_{4}$&$C_{2h}$&Screw \cite{Clark2021}&(0,0,1)&(0,0,1)&15&(0,0,1)\\
		11. RbAg$_{2}$Fe(VO$_{4}$)$_{2}$&$C_{3i}$&Screw \cite{Amuneke2014}&(0,0,1)&(0,0,1)&3.2&(0,0,1)\\
		12. KAg$_{2}$Fe(VO$_{4}$)$_{2}$&$C_{3i}$&Screw \cite{Amuneke2014}&(0,0,1)&(0,0,1)&2.8&(0,0,1)\\
		\hline
	\end{tabular}
\end{table*}

\textit{Applications--}To the best of our knowledge, 32 type-II multiferroic materials exhibiting incommensurate magnetic order have been experimentally confirmed to date. Their point groups, magnetic orders, propagation vectors, spin spiral axes, and polarization axes are listed in Table S1 of the supplementary materials (SM). Our theory successfully explains the magnetic-order-induced polarizations in all these materials, as shown in section II of the SM. According to distinct magnetic configurations: (a) cycloidal order and (b) screwed order, we take two typical cases to elaborate MEC effects in detail next.

The first representative material is rare earth multiferroic TbMnO$_{3}$, which is crystallized in group $D_{2h}$ rather than group $O_{h}$ \cite{PhysRevLett.95.087206}. Symmetry analysis reveals that the free energy density of MEC terms for group $D_{2h}$ adopts the following explicit form:
\begin{equation}\begin{array}{ccccccc}
		\Phi_{em}(\mathbf P,\mathbf M)=\alpha_{1}P_{x}(M_{x}\nabla_{y}M_{y}-M_{y}\nabla_{y}M_{x})\\
		+\alpha_{2}P_{x}(M_{x}\nabla_{z}M_{z}-M_{z}\nabla_{z}M_{x})\\
		+\alpha_{3}P_{y}(M_{y}\nabla_{z}M_{z}-M_{z}\nabla_{z}M_{y})\\
		+\alpha_{4}P_{y}(M_{y}\nabla_{x}M_{x}-M_{x}\nabla_{x}M_{y})\\
		+\alpha_{5}P_{z}(M_{z}\nabla_{x}M_{x}-M_{x}\nabla_{x}M_{z})\\
		+\alpha_{6}P_{z}(M_{z}\nabla_{y}M_{y}-M_{y}\nabla_{y}M_{z}),
\end{array}\end{equation}
which structurally resembles the antisymmetric Lifshitz invariant. By performing a variational derivative of the total free energy $\Phi_{0}+\Phi_{em}$ with respect to $\mathbf P$, where the polarization energy $\Phi_{0}(\mathbf P)=\frac{\mathbf P^{2}}{2\chi}$ incorporates the dielectric susceptibility $\chi$, we get the final results as follows
\begin{equation}\begin{array}{cccccc}P_{x}=\alpha_{1}\chi(M_{x}\nabla_{y}M_{y}-M_{y}\nabla_{y}M_{x})\\
		+\alpha_{2}\chi(M_{x}\nabla_{z}M_{z}-M_{z}\nabla_{z}M_{x}),\\
		P_{y}=\alpha_{3}\chi(M_{y}\nabla_{z}M_{z}-M_{z}\nabla_{z}M_{y})\\
		+\alpha_{4}\chi(M_{y}\nabla_{x}M_{x}-M_{x}\nabla_{x}M_{y}),\\
		P_{z}=\alpha_{5}\chi(M_{z}\nabla_{x}M_{x}-M_{x}\nabla_{x}M_{z})\\
		+\alpha_{6}\chi(M_{z}\nabla_{y}M_{y}-M_{y}\nabla_{y}M_{z}).\end{array}\end{equation}
The complete set of MEC terms for the remaining point groups is systematically listed in section I of the SM. Below the magnetic transition temperature $T_{M}=28$ K, the polarization is induced by the incommensurate cycloidal magnetic structure in TbMnO$_{3}$ along the $z$ direction \cite{PhysRevLett.92.257201}, as shown in Table I. Considering a spiral magnetic structure given as follows:
\begin{equation}\mathbf M=M\cos(\mathbf q\cdot\mathbf r+\phi_{0})\mathbf e_{1}+M\sin(\mathbf q\cdot\mathbf r+\phi_{0})\mathbf e_{2},\end{equation}
where $\phi_{0}$ is an initial phase, $\mathbf q$ represents the propagation vector, and $\mathbf e$ represents unit vector. The cycloidal magnetic structure [Fig. 1 (a)] in TbMnO$_{3}$ is given by $\mathbf q\parallel$ $y$ and the spin spiral axis $(\mathbf e_{1}\times\mathbf e_{2})\parallel$ $x$ in the Cartesian coordinate system (see SM). Taking Eq. (3) into Eq. (2), we find that the average polarization $\bar{\mathbf P}=\frac{1}{V}\int \mathbf P dV$ in TbMnO$_{3}$ is given by
\begin{equation}\begin{array}{llllll}\bar P_{x}=\bar P_{y}=0,
		\bar P_{z}=-\alpha_{6}\chi M^{2}q\end{array}\end{equation}
oriented along the $z$ direction, which agrees with the experimental results \cite{PhysRevLett.92.257201}. Due to the spin-cycloidal order, the polarization expressions $\bar P_{x}=\bar P_{y}=0, \bar P_{z}=-\alpha_{0}\chi M^{2}q$ calculated from the octahedral group $O_{h}$ for TbMnO$_{3}$ show fortuitous agreement with the experimental results (see Table I).

The second representative material is magnet Cu$_{2}$OSeO$_{3}$ \cite{PhysRevLett.128.037201,Mochizuki+2020,PhysRevLett.108.237204,Ruff2015} crystallizes in the noncentrosymmetric group $T$ and hosts an incommensurate screwed magnetic ground state characterized by spin spiral axis parallel to the magnetic propagation vector $(\mathbf e_{1}\times\mathbf e_{2})\parallel\mathbf q$ [Fig. 1 (b)]. There are several different ferroelectric polarization phases observed \cite{PhysRevB.86.060403}: (i) $\mathbf P\parallel$ [111] when $\mathbf q\parallel$ [111], (ii) $\mathbf P\parallel$ [001] when $\mathbf q\parallel$ [110], and (iii) $\mathbf P=0$ when $\mathbf q\parallel$ [001]. Following the previous steps, we get the MEC terms linear in the first derivatives of $\mathbf M$ for group $T$ as 
\begin{equation}\begin{array}{llllll}P_{x}=\\
		\alpha_{1}\chi(M_{x}\nabla_{y}M_{y}-M_{y}\nabla_{y}M_{x})+\alpha_{2}\chi(M_{x}\nabla_{z}M_{z}-M_{z}\nabla_{z}M_{x}),\\
		P_{y}=\\
		\alpha_{1}\chi(M_{y}\nabla_{z}M_{z}-M_{z}\nabla_{z}M_{y})+\alpha_{2}\chi(M_{y}\nabla_{x}M_{x}-M_{x}\nabla_{x}M_{y}),\\
		P_{z}=\\
		\alpha_{1}\chi(M_{z}\nabla_{x}M_{x}-M_{x}\nabla_{x}M_{z})+\alpha_{2}\chi(M_{z}\nabla_{y}M_{y}-M_{y}\nabla_{y}M_{z}).\end{array}\end{equation}
It is tedious and lengthy to show the MEC terms quadratic in the first derivatives of $\mathbf M$, so we derive them in SM. Taking Eq. (3) into Eq. (5) and Eqs. (16-18) from SM, we get the total polarization components as:
\begin{equation}\begin{array}{llllll}\bar P_{x}=\bar P_{y}=\bar P_{z}=(\alpha_{1}-\alpha_{2})\chi M^{2}q/3\\+(2\sum_{i=1\sim3}\beta_{i}-\sum_{j=4\sim8}\beta_{j})\chi M^{2}q^{2}/9\end{array}\end{equation}
for $\mathbf q\parallel$ [111],
\begin{equation}\begin{array}{llllll}\bar P_{x}=\bar P_{y}=0$, $\bar P_{z}=(\alpha_{2}-\alpha_{1})\chi M^{2}q/2\\+(2\beta_{1}+\beta_{2}+\beta_{3}-\beta_{7}-\beta_{8})\chi M^{2}q^{2}/6\end{array}\end{equation}
for $\mathbf q\parallel$ [110], and complete polarization suppression $\bar{\mathbf P}=0$ for $\mathbf q\parallel$ [001], showing full agreement with experimental measurements. Due to the inversion asymmetry of group $T$, both MEC terms linear and quadratic in the first derivatives of $\mathbf M$ make contributions to the polarization in Cu$_{2}$OSeO$_{3}$. From Table I, our theoretical framework successfully accounts for the MEC effects in type-II multiferroics with spin-screwed orders, whereas Mostovoy's study fails to explain.
\begin{figure*}
	\centering
	\includegraphics[width=15.4cm]{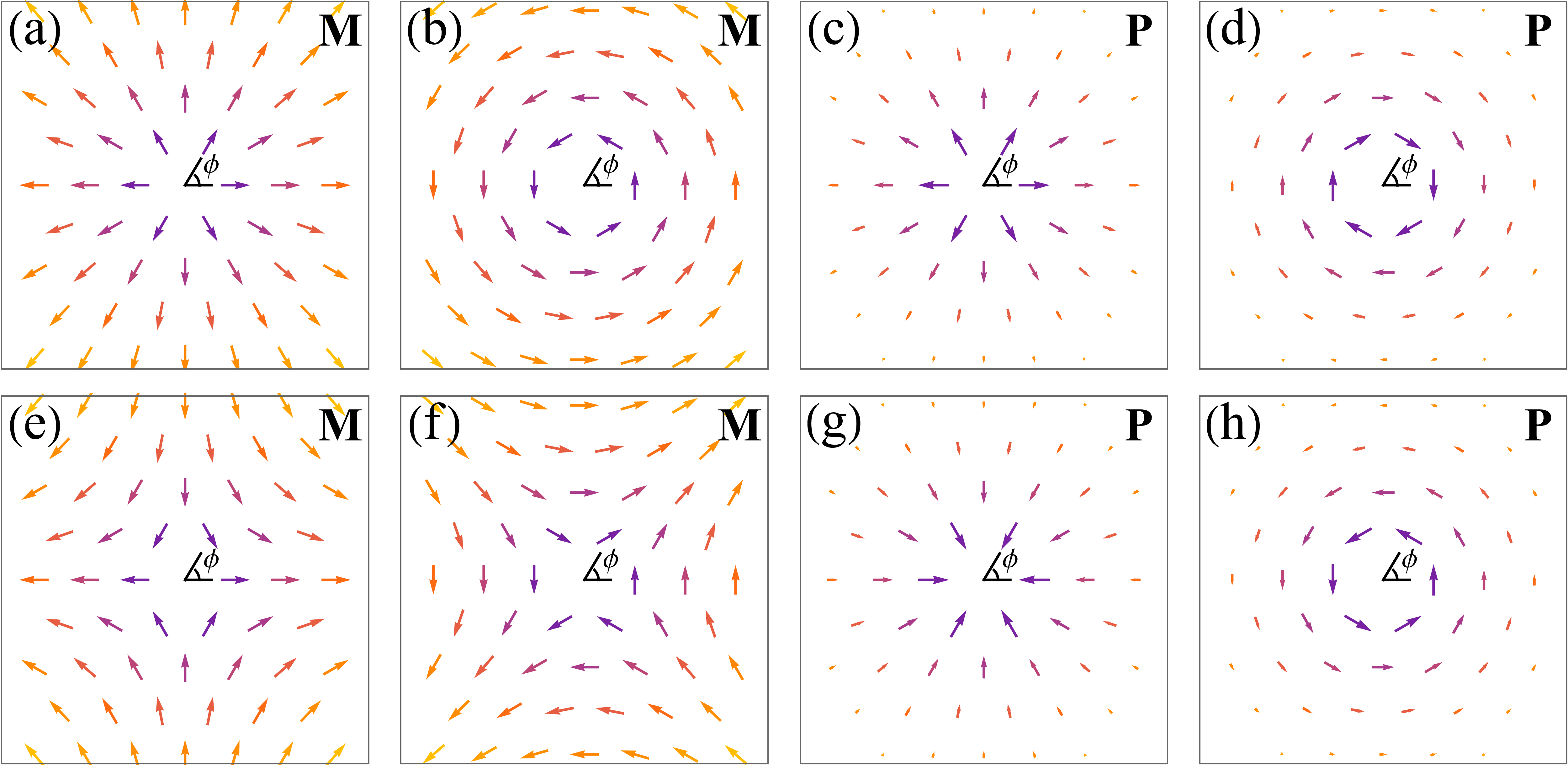}
	\caption{Schematic illustrations of topological ferromagnetic states and induced ferroelectric states. Both (a) the ferromagnetic monopole ($n=1$, $\phi_{0}=0$) and (b) the ferromagnetic vortex ($n=1$, $\phi_{0}=\pi/2$) can generate (c) the ferroelectric monopole and (d) the ferroelectric vortex. Both (e) the ferromagnetic antivortex ($n=-1$, $\phi_{0}=0$) and (f) the ferromagnetic antivortex ($n=-1$, $\phi_{0}=\pi/2$) can generate (g) the opposite ferroelectric monopole and (h) the ferroelectric vortex.}\label{fig1}
\end{figure*}

\textit{Predictions--}Our theoretical framework has been tested against available data, providing a basis for further predictions. To explore potential candidate materials, we conduct a systematic survey of the MAGNDATA database \cite{Gallego2016,https://doi.org/10.1107/S1600576716015491} in which the magnetic orders, the magnetic propagation vectors, and the transition temperatures have been experimentally well-established. The MAGNDATA database comprehensively documents 2317 magnetic materials, including 2156 commensurate structures and 161 incommensurate structures. Our theory predicts 12 insulating incommensurate structures to be type-II multiferroics, excluding metallic compounds. Remarkably, the highest reported magnetic transition temperature among these compounds reaches 84 K. Furthermore, we calculate the directions of magnetically induced ferroelectric polarization for each of these materials, and the results are completely summarized in Table II. Interestingly, we find that the type-I multiferroics can simultaneously exhibit the characteristics of type-II multiferroics. Specifically, magnetic materials LiFeAs$_2$O$_7$ \cite{PhysRevB.88.214433}, Ni$_2$InSbO$_6$ \cite{Ivanov2013}, and Ni$_2$ScSbO$_6$ \cite{Ivanov2013} possess intrinsic ferroelectric polarization, and thus are conventionally classified as type-I multiferroics since the polarization and the magnetization arise from different mechanisms. However, our theoretical analysis reveals that magnetic orders in these materials induce additional polarization components oriented differently from the intrinsic polarization. Taking Ni$_{2}$InSbO$_{6}$ as an illustrative example (more details provided in SM), we note that it crystallizes in polar point group $C_{3}$ with screwed magnetic structure characterized by wave vector $\mathbf q\parallel$ $y$. Taking Eq. (3) into Eq. (6) and Eqs. (40-42) from SM, we get the average polarization
\begin{equation}\begin{array}{llllll}
		P_{x}=-\alpha_{8}\chi M^{2}q+(\beta_{7}-\beta_{21})M^{2}\chi q^{2},\\
		P_{y}=-\alpha_{7}\chi M^{2}q+(\beta_{10}-\beta_{23})M^{2}\chi q^{2},\\
		P_{z}=-\alpha_{6}\chi M^{2}q+(\beta_{30}+\beta_{36})M^{2}\chi q^{2}.\end{array}\end{equation}
One can see that the magnetic-order-induced polarization does not align with the intrinsic polarization direction $z$ axis. Such an orientation discrepancy is experimentally measurable and offers a robust method to verify the coexistence of type-I and type-II multiferroic behaviors within a single compound.

In addition, our theory can be applied to predict novel MEC patterns. According to the quadratic MEC effect derived in subsection B of section I of SM, we find that even collinear sinusoidal spin configurations: $\mathbf M=M\cos(\mathbf q\cdot\mathbf r)\mathbf e$ can introduce nonzero ferroelectric polarization $\mathbf P\neq0$ [Fig. 1 (c)], besides the spin-cycloidal and the spin-screw orders mentioned before. Nonetheless, it seems that collinear magnets with sinusoidal magnetic order have not yet been reported as ferroelectric materials \cite{PhysRevB.71.224425,PhysRevB.79.134426,PhysRevB.63.094411,PhysRevB.76.104405,PhysRevB.48.6087,PhysRevB.90.024408,PhysRevB.87.144403,PhysRevB.81.214407,PhysRevB.83.134410}. From a symmetry perspective, this absence may result from the preservation of SIS in these systems. Since the sinusoidal magnetic order alone does not break SIS, the emergence of ferroelectric polarization is consequently prohibited.

\textit{Ferroic vortex states--}The emergence of ferroelectric vortex states introduced by ferromagnetic vortex states is predicted by our theory as well. The in-plane rotation combined with time-reversal symmetry, such as $C_{2}T$, preserves ferromagnetic vortices but breaks ferroelectric vortices. Thus, only 11 uniaxial point groups: $C_{6h}$, $C_{3h}$, $C_{6}$, $C_{4h}$, $S_{4}$, $C_{4}$, $C_{3i}$, $C_{3}$, $C_{2h}$, $C_{s}$, and $C_{2}$, are simultaneously compatible with both ferromagnetic and ferroelectric vortices. For a magnetic vortex (see Fig. 2) described by
\begin{equation}\mathbf M=M\cos(n\phi+\phi_{0})\mathbf e_{1}+M\sin(n\phi+\phi_{0})\mathbf e_{2},\end{equation}
where $n$ is the winding number and $\phi$ represents the azimuthal angle, the induced polarization in these uniaxial point groups [using Eq. (4), (6), and (8) in SM] is given by
\begin{equation}\mathbf P=n\chi M^{2}[\alpha_{1}(\cos\phi \mathbf e_{1}+\sin\phi \mathbf e_{2})+\alpha_{4}(\sin\phi \mathbf e_{1}-\cos\phi \mathbf e_{2})]/r.\end{equation}
The first term corresponds to the ferroelectric monopole [Fig. 2 (c) and (g)] with quantized charge localized at the vortex core mentioned in ref. \cite{PhysRevLett.96.067601}, while the second term represents the ferroelectric vortex state with winding number $\pm$1 [Fig. 2 (d) and (h)] driven by the ferromagnetic vortex state. Thus, this study paves the way for realizing the coexistence of ferromagnetic and ferroelectric vortex states in multiferroic materials with uniaxial space groups No. 3-15, No. 75-88, No. 143-148, and No.168-176. 

\textit{Conclusions--}In conclusion, we establish a comprehensive symmetry-based framework for MEC effects spanning all 32 point groups through a phenomenological Landau theory. The MEC effects in all type-II multiferroics with incommensurate magnetic structures can be explained properly by our theory. Moreover, we predict 12 candidate type-II multiferroics with a maximum magnetic transition temperature of upto 84 K. We find that the type-I multiferroics can simultaneously exhibit the characteristics of type-II multiferroics. Theoretically, we also demonstrate that not only conventional cycloidal and screwed magnetic orders but also sinusoidal magnetic order can generate ferroelectric polarization. Finally, we propose that the MEC is a feasible approach to realize the coexistence of the ferromagnetic and the ferroelectric vortex states, effectively circumventing the conventional limitations imposed by the scale mismatch. Notably, our theory is also applicable for two-dimensional materials.

\textit{Acknowledgments--}This work was financially supported by the National Natural Science Foundation of China (Grants No. 12188101, No. 12334007, and No. 12474233), the National Key R\&D Program of China (Grant No. 2022YFA1403601), Innovation Program for Quantum Science and Technology (Grant No. 2021ZD0301902, and No. 2024ZD0300101), Natural Science Foundation of Jiangsu Province (No. BK20233001, and No. BK20243011), and China Postdoctoral Science Foundation (2023M731615).

	\bibliography{bibl}
	
\end{document}